\documentclass[aps,prb,twocolumn,showpacs,superscriptaddress,10pt, citeautoscript]{revtex4-1}

\usepackage[utf8]{inputenc}
\usepackage[english]{babel}

\usepackage{graphicx}
\usepackage{amsmath,amssymb,amstext,amsthm}
\usepackage{bbold}
\usepackage{cancel}
\usepackage{color}
\usepackage{verbatim}
\usepackage{booktabs}

\newcommand{\Z}{\mathbb{Z}}

\DeclareUnicodeCharacter{2032}{$^\prime$}

\usepackage[%
  colorlinks=true,
  urlcolor=blue,
  linkcolor=blue,
  citecolor=blue
]{hyperref}

\begin{document}

\title{The influence of lattice termination on the edge states of the\\quantum spin Hall insulator monolayer $1T'$-WTe$_2$}

\author{Alexander Lau}
\affiliation{Kavli Institute of Nanoscience, Delft University of Technology, P.O. Box 4056, 2600 GA Delft, Netherlands}

\author{Rajyavardhan Ray}
\affiliation{Institute for Theoretical Solid State Physics, IFW Dresden, Helmholtzstr. 20, 01069 Dresden, Germany}
\affiliation{Dresden Center for Computational Materials Science (DCMS), TU Dresden, 01062 Dresden, Germany}

\author{D{\'a}niel Varjas}
\affiliation{Kavli Institute of Nanoscience, Delft University of Technology, P.O. Box 4056, 2600 GA Delft, Netherlands}
\affiliation{QuTech, Delft University of Technology, P.O. Box 4056, 2600 GA Delft, The Netherlands}

\author{Anton Akhmerov}
\affiliation{Kavli Institute of Nanoscience, Delft University of Technology, P.O. Box 4056, 2600 GA Delft, Netherlands}

\date{\today}

\begin{abstract}
We study the influence of sample termination on the electronic properties of the novel quantum spin Hall insulator monolayer $1T'$-WTe$_2$.
For this purpose, we construct an accurate, minimal 4-orbital tight-binding model with spin-orbit coupling by employing a combination of density-functional theory calculations, symmetry considerations, and fitting to experimental data.
Based on this model, we compute energy bands and 2-terminal conductance spectra for various ribbon geometries with different terminations, with and without magnetic field.
Because of the strong electron-hole asymmetry we find that the edge Dirac point is buried in the bulk bands for most edge terminations.
In the presence of a magnetic field, an in-gap edge Dirac point leads to exponential suppression of conductance as an edge Zeeman gap opens, whereas the conductance stays at the quantized value when the Dirac point is buried in the bulk bands.
Finally, we find that disorder in the edge termination drastically changes this picture: the conductance of a sufficiently rough edge is uniformly suppressed for all energies in the bulk gap regardless of the orientation of the edge.
\end{abstract}

\maketitle

\section{Introduction}

Quantum spin Hall (QSH) insulators are two-dimensional (2D) materials
with a pair of counter-propagating, helical electronic modes along their edges  protected from backscattering by time-reversal symmetry~\cite{KaM05_1,KaM05_2,BeZ06,SWS06,FuH07,KWB07}.
They represent celebrated examples of topological insulators~\cite{HaK10,QiZ11,RSF10,Lud15} and have a nonzero $\Z_2$ topological invariant $\nu$.
Their hallmark is a quantized electronic conductance of $G = 2 e^2/h$ as a consequence of the eponymous quantum spin Hall effect.
More specifically, each of the two oppositely spin-polarized, conducting edge channels contributes one conductance quantum $e^2/h$.
This is a robust feature since backscattering between the two edge channels is forbidden by time-reversal symmetry.
The described robustness, however, is expected to break down once time-reversal symmetry is broken, for instance by a magnetic field.

Experimental efforts in realizing QSH insulators have mostly focused on quantum-well heterostructures based on three-dimensional semiconductors such as HgTe/CdTe or InAs/GaSb quantum wells~\cite{KWB07,KDS11,PiH14,DKS15,MCW15,NSK16}.
Recently, technological advances in the preparation of monolayer materials have shifted the focus to truly 2D materials as platform for QSH insulators~\cite{QLF14,ZFL14,Eza15,LJZ16}.
Those materials include the graphene descendants silicene, germanene, and stanene~\cite{LFY11,XYZ13,Eza15}, as well as the related Bismuthene~\cite{LHH18}.

Another materials class that has attracted attention in this context are single-layer transition metal dichalcogenides~\cite{QLF14,LeS17,UPT18,CPC18,XMS18,ZBF18,ShS18,FWC18,SPF18,KLS18,WRS18}. Specifically, monolayer WTe$_2$ has been identified theoretically to realize a QSH phase~\cite{QLF14,CSC16,XXL16,ZCG16}, and various experiments have verified its key features: an insulating bulk, the presence of conducting edge channels, and a quantized electronic conductance of $2e^2/h$ over a large range of temperatures~\cite{FPW17,JSL17,PYL17,SJZ17,TZW17,SKJ18,WFG18}.
Furthermore, a magnetic field has been shown to lead to a breakdown of conductance in the investigated samples~\cite{WFG18}.
This suggests the opening of a Zeeman-type gap in the edge-state spectrum, contrary to other QSH insulator candidate materials~\cite{MCW15,NSK16,SPA18}.
It remains elusive, however, whether the presence of a Zeeman gap is a generic feature of single-layer WTe$_2$, or which type of edge terminations show such a gap.
To answer this question on a theoretical footing, an accurate low-energy lattice model is required. 

In this article, we study the electronic properties of monolayer WTe$_2$ in the framework of a minimal 4-orbital tight-binding model with spin-orbit coupling (SOC).
In order to obtain an accurate low-energy description, we derive our model using a hybrid approach that combines density-functional theory (DFT) calculations, symmetry considerations, and experimental data from angle-resolved photoemission spectroscopy (ARPES).
We first show that our model captures all the essential features that contribute to the nontrivial topology of the material.
When applied to different nanoribbon geometries, this model provides a qualitative understanding of the influence of a magnetic field and of edge disorder:
we find that the exponential suppression of conductance, associated with an edge Zeeman gap, is observed only for sufficiently clean edges that feature an in-gap edge Dirac point.

\section{Tight-binding model for monolayer WTe$_2$}

WTe$_2$ is a member of the transition metal dichalcogenide family of materials. Single layers in this materials class crystallize in a variety of polytypic structures such as the hexagonal $2H$, the tetragonal $1T$, and the distorted $1T'$ structure~\cite{HeK99,EFY12,QLF14}.
While the former represent trivial semiconductors, the $1T'$ configuration gives rise to 2D QSH insulators~\cite{QLF14}.
Furthermore, monolayer WTe$_2$ is the only member of this materials class that is known to realize the topologically nontrivial $1T'$ structure as its stable ground state~\cite{QLF14,DLR14,UPT18,CPC18}.

\begin{figure}[t]\centering
\includegraphics[width=1.0\columnwidth]
{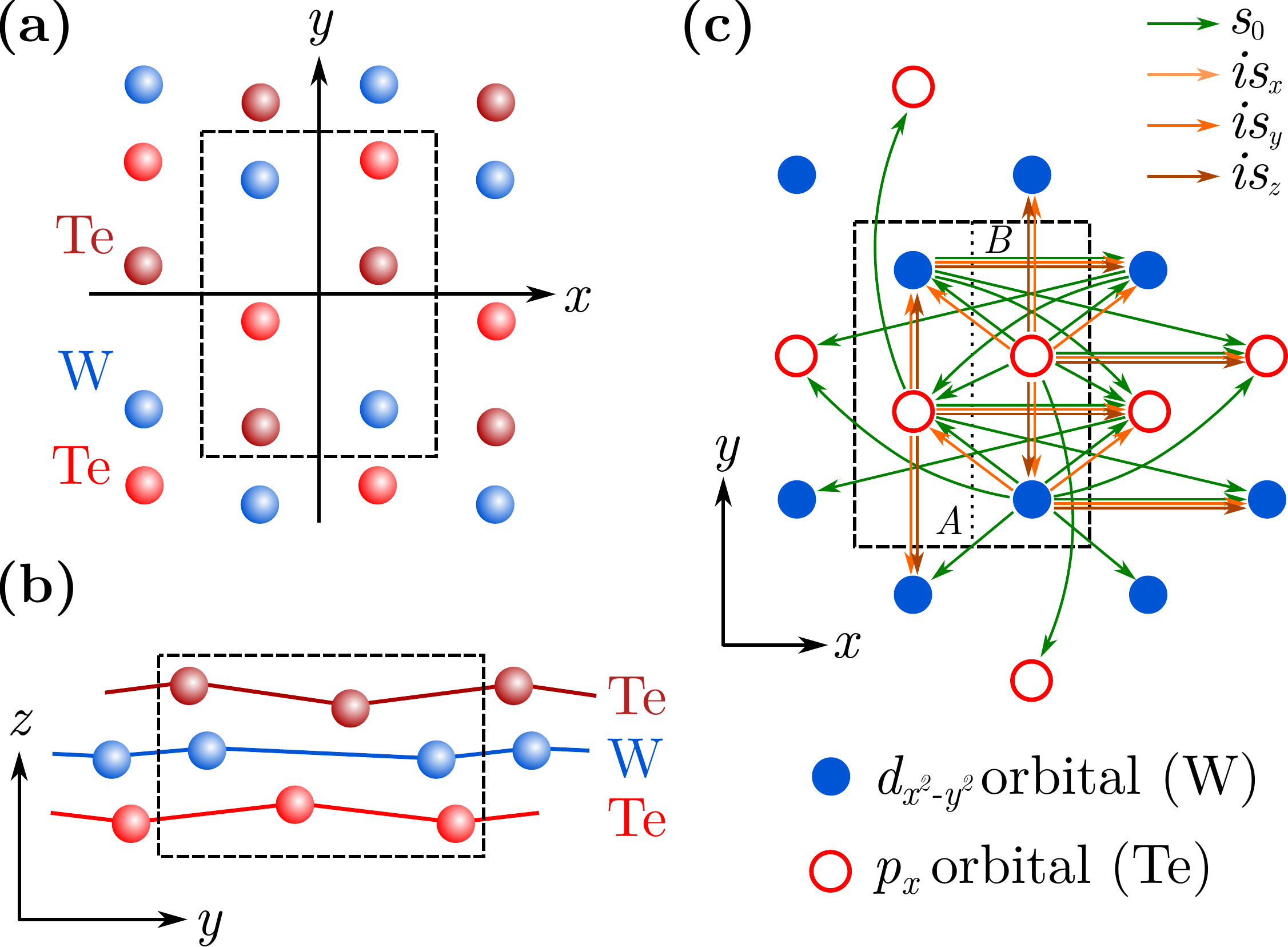}
\caption{
Lattice and tight-binding model of WTe$_2$: (a) top view of the lattice; (b) side view of the lattice. The dashed box indicates the 6-site unit cell. (c) Reduced 4-site lattice used for the tight-binding model with sublattices A and B indicated. The arrows depict the hopping terms ($s_0$) and the SOC terms ($s_{x,y,z}$) of the model. We provide a more detailed schematic in the Supplemental Material~\cite{supp}.
}
\label{fig:WTe2_unit_cells}
\end{figure}

The crystal structure is generated from the tetragonal $1T$ configuration by a lattice distortion~\cite{HeK99,EFY12,QLF14}: the original $1T$ configuration is composed of three hexagonal layers in rhombohedral ABC stacking with one W layer sandwiched between two Te layers.
A structural distortion then shifts the W atoms along the $b$ direction to form zigzag chains.
At the same time, the three atomic sheets of the lattice are buckled in the out-of-plane direction. The resulting $1T'$ structure has a rectangular unit cell, containing two W sites and four Te sites.
We show a top view and a side view of the lattice in Figs.~\ref{fig:WTe2_unit_cells}(a) and~(b), respectively.

The lattice distortion of the $1T'$ configuration is the key ingredient to the nontrivial topology~\cite{QLF14,CSC16} of WTe$_2$. It enlarges the unit cell and leads to a reconstruction of the electronic bands in the 2D Brillouin zone.
As a consequence, a band inversion takes place at the $\Gamma$ point~\cite{QLF14,CSC16}. In contrast to other QSH insulator materials, this band inversion is \emph{not} induced by SOC. In fact, theoretical calculations show that the respective bands are inverted already in the nonrelativistic limit, i.e., when SOC is neglected.
The corresponding ``spinless'' band structure, however, does not have a full bulk energy gap.
Instead, it realizes a 2D type-II Dirac semimetal due to the presence of two tilted, unpinned Dirac cones near the $\Gamma$ point, protected by a glide reflection symmetry~\cite{MAN16}.
Recovering the spin-degree of freedom by including SOC then lifts the degeneracy at the Dirac cones and leads to a sizeable, indirect bulk energy gap.

Our aim is to investigate the electronic properties of WTe$_2$ in ribbons with arbitrary terminations including rough edges.
So far, theoretical studies of this material have mostly been limited to DFT calculations and low-energy $\mathbf{k}\cdot\mathbf{p}$ models~\cite{QLF14,ZCG16,XXL16,CSC16,LeS17,ZBF18,ShS18}.
For the former, it is computationally expensive to model large ribbons with disordered edges.
The latter, on the other hand, is a long wave-length theory and does not capture geometric details at small length scales.
In the following, we therefore construct a minimal tight-binding model suitable for studying the low-energy electronic properties of monolayer WTe$_2$ on arbitrarily terminated lattices.

Effective tight-binding models for monolayer WTe$_2$ have been derived from DFT calculations in Refs.~\onlinecite{MAN16,OMD18}.
Such models are typically extracted from DFT via a projection of the DFT wavefunctions onto a subset of atomic orbitals.
Fully-relativistic DFT calculations, which take into account the spin degree of freedom of the electrons, however,  have not been able to reproduce quantitatively the size of the energy gap and the dispersion of the valence band close to the Fermi level, as measured in scanning tunneling spectroscopy and ARPES experiments~\cite{ZCG16,TZW17,JSL17,CGC18}.
To overcome this issue, we choose to take an alternative route by combining scalar-relativistic DFT calculations with symmetry considerations and data from ARPES, thereby obtaining an accurate bulk model.

\subsection{DFT calculations and optimized Wannier fit}

As a first step of our construction, we perform DFT calculations of the electronic structure of freestanding $1T'$-WTe$_2$ monolayers in an all-electron full-potential local-orbital basis using the FPLO code~\cite{KoE99, fplo_web}.
We employ the PBE implementation~\cite{PBE96} of the generalized gradient
approximation (GGA) including scalar relativistic corrections.
For the numerical integration, we use a ($12 \times 12 \times 1$) $\mathbf{k}$-mesh in the full Brillouin zone along with the linear tetrahedron method.
Moreover, we take the lattice parameters and the atomic positions from Ref.~\onlinecite{MJI92}.

\begin{figure}[t]\centering
\includegraphics[width=1.0\columnwidth]
{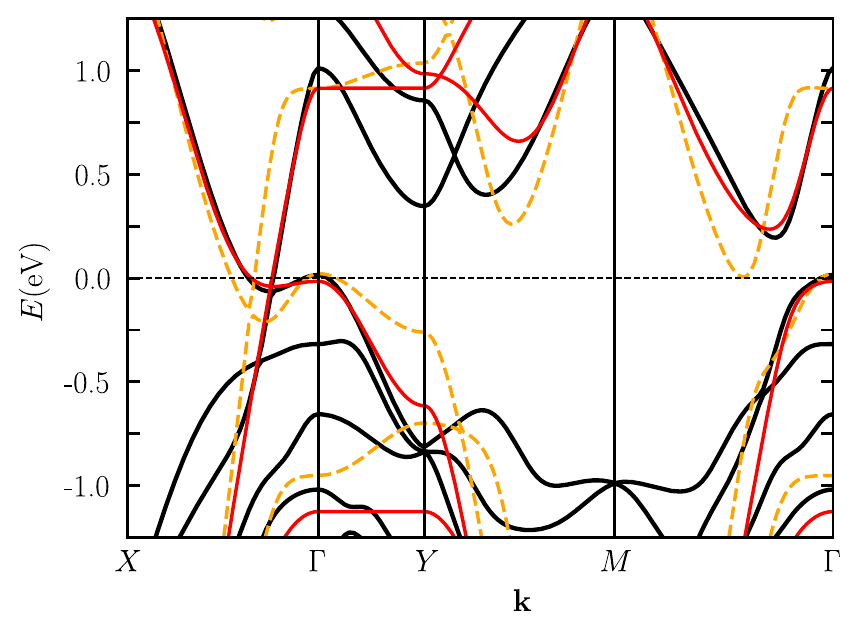}
\caption{Energy bands without SOC along high-symmetry lines in the Brillouin zone: DFT calculations (solid black lines), Wannier tight-binding fit (dashed orange lines), and energy bands from the optimized tight-binding fit (solid red lines).}
\label{fig:DFT_and_tight_binding}
\end{figure}

In agreement with the literature~\cite{MAN16,CSC16}, we find that the four bands closest to the Fermi level are composed mainly of contributions from two $3d_{x^2-y^2}$-type orbitals centered at W sites and from two $5p_x$-type orbitals centered at a subset of Te sites.
Hence, a tight-binding model consisting of these four orbitals represents a suitable minimal model. We therefore construct localized Wannier orbitals from the two W-$d$ and two Te-$p$ orbitals [see Fig.~\ref{fig:WTe2_unit_cells}(c)], and compute the corresponding real-space overlaps to obtain the tight-binding parameters for our minimal 4-orbital model. Figure~\ref{fig:DFT_and_tight_binding} shows the resulting tight-binding bands, corresponding to the Wannier fit, along with the DFT band structure.

The Wannier tight-binding model reproduces the key features of the DFT band structure close to the Fermi level reasonably well, including the tilted type-II Dirac cones near the $\Gamma$ point.
It turns out, however, that the small number of orbitals requires the inclusion of a large number of neighbors in the Wannier fit.
At the same time, irrespective of the number of neighbors, the location of the Dirac point in $\mathbf{k}$ space is slightly shifted compared to the DFT bandstructure.
To obtain a more accurate low-energy description, we therefore adopt a hybrid approach: we take the onsite energies and the largest hopping parameters ($\geq 0.4\mathrm{eV}$) from DFT, while we treat others above a certain energy cutoff as effective hopping parameters (see Supplemental Material for details~\cite{supp}).
Finally, we optimize the latter to fit the DFT energy bands close to the Fermi level.
In the Supplemental Material~\cite{supp}, we demonstrate that such an optimization is indeed needed in order not to overestimate the SOC terms in the next step and that this does not affect the final conclusions of our study.

Figure~\ref{fig:WTe2_unit_cells}(c) illustrates the hopping terms of the resulting tight-binding Hamiltonian without SOC.
We provide a more detailed depiction in the Supplemental Material~\cite{supp}.
The corresponding Bloch Hamiltonian is
\begin{eqnarray}
H_0  &=&
\Big[\frac{\mu_p}{2} + t_{px} \cos(a k_x) + t_{py} \cos(b k_y)\Big]\, \Gamma_1^- \nonumber\\
&& + \Big[\frac{\mu_d}{2} + t_{dx} \cos(a k_x)\Big]\, \Gamma_1^+ \nonumber\\
&& +\, t_{dAB}\, e^{-ibk_y} \big(1 + e^{iak_x}\big) e^{i\mathbf{k}\cdot\Delta_1} \Gamma_2^+ \nonumber\\
&& +\, t_{pAB}\, \big(1 + e^{iak_x}\big) e^{i\mathbf{k}\cdot\Delta_2} \Gamma_2^- \nonumber\\
&& +\, t_{0AB}\, \big(1 - e^{iak_x}\big) e^{i\mathbf{k}\cdot\Delta_3} \Gamma_3 \nonumber\\
&& -\, 2i t_{0x}\, \sin(a k_x)
\big[ e^{i\mathbf{k}\cdot\Delta_4} \Gamma_4^+
+ e^{-i\mathbf{k}\cdot\Delta_4} \Gamma_4^-\big] \nonumber\\
&& +\, t_{0ABx}\, \big(e^{-iak_x} - e^{2iak_x}\big) e^{i\mathbf{k}\cdot\Delta_3} \Gamma_3 \nonumber\\
&& +\, \mathrm{H.c.}
\label{eq:spinless_Hamiltonian}
\end{eqnarray}
The $4\times 4$ matrices $\Gamma_i$ are linear combinations of products $\tau_j\sigma_i$ of Pauli matrices (see Supplemental Material~\cite{supp}) acting in orbital space with respect to the basis $\lbrace d_{\mathbf{k}Ads}, d_{\mathbf{k}Aps}, d_{\mathbf{k}Bds}, d_{\mathbf{k}Bps}\rbrace$, where $d_{\mathbf{k}cls}$ annihilates an electron with momentum $\mathbf{k}$, spin-$S_z$ eigenvalue $s=\uparrow,\downarrow$ and orbital $l=p,d$ (Te, W) in sublattice $c=A,B$. We have further defined $\Delta_1 = \mathbf{r}_{Ad}-\mathbf{r}_{Bd}$, $\Delta_2 = \mathbf{r}_{Ap}-\mathbf{r}_{Bp}$, $\Delta_3 = \mathbf{r}_{Ad}-\mathbf{r}_{Bp}$ and $\Delta_4 = \mathbf{r}_{Ad}-\mathbf{r}_{Ap}$, where the vector $\mathbf{r}_{cl}$ denotes the position of the corresponding lattice site, associated with an orbital $l$ in sublattice $c$, in the unit cell.
The lattice constants are  $a=3.477\,\mbox{\AA}$ and $b=6.249\,\mbox{\AA}$.

\begin{figure}[t]\centering
\includegraphics[width=1.0\columnwidth]
{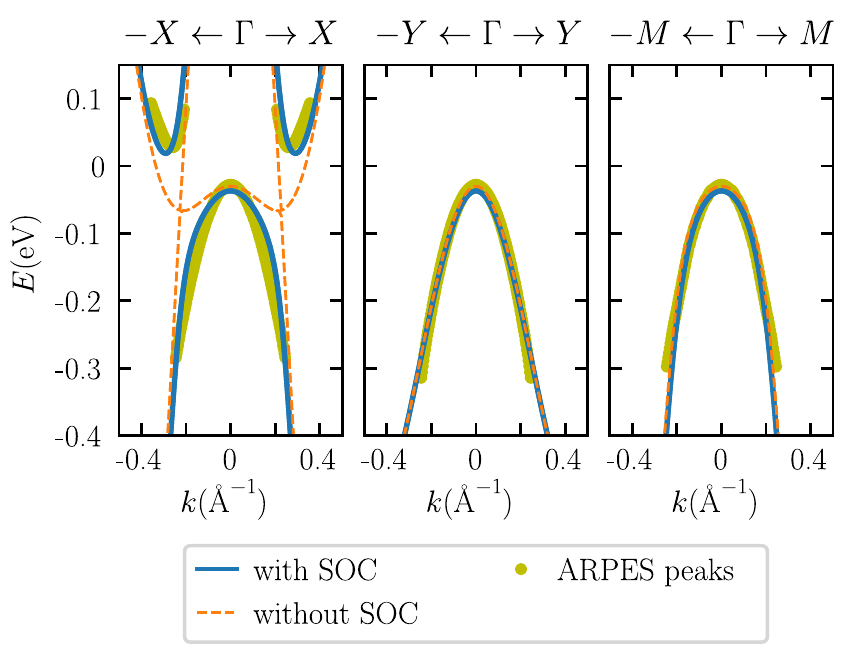}
\caption{Energy bands of the final tight-binding model with and without SOC ($\lambda_i$=0) compared with intensity peaks from ARPES measurements (courtesy of Ref.~\onlinecite{TZW17}).}
\label{fig:SOC_fit}
\end{figure}
Figure~\ref{fig:DFT_and_tight_binding} shows a comparison of the energy bands of the optimized tight-binding model with the DFT bands close to the Fermi level.
We find that the model indeed gives an accurate description of the DFT bands around the Fermi level and correctly reproduces the positions of the Dirac points.

\subsection{Spin-orbit coupling from symmetries}

Having established an accurate minimal tight-binding model without SOC, we now add a set of symmetry-allowed SOC terms to our model.
The symmetry group of the WTe$_2$ lattice is generated by the following operations~\cite{MAN16}: lattice translations $T(\mathbf{a})$ and $T(\mathbf{b})$, where $\mathbf{a}=a\hat{\mathbf{x}}$ and $\mathbf{b}=b\hat{\mathbf{y}}$ with unit vectors $\hat{\mathbf{x}}$, $\hat{\mathbf{y}}$; a glide reflection $\tilde{M}_x = T(\mathbf{a}/2)M_x$ with the regular reflection $M_x$ about the $yz$ plane acting as $M_x:(x,y,z)\rightarrow(-x,y,z)$; inversion $\mathcal{I}$ acting as $\mathcal{I}:(x,y,z)\rightarrow(-x,-y,-z)$; and time reversal $\Theta$.
In the basis of the Hamiltonian in Eq.~\eqref{eq:spinless_Hamiltonian}, the matrix representations of the symmetry operations are $\hat{\Theta}=is_y\Gamma_0 K$ with complex conjugation $K$, $\hat{\mathcal{I}}=s_0\tau_1\sigma_3$, and $\hat{M}_x = is_x\tau_0\sigma_3$ for the non-translational component of $\tilde{M}_x$.

\begin{table}[t]
\setlength{\tabcolsep}{7pt}
\begin{tabular}{|c|r||c|r|}
\hline
$\mu_p$ & $-1.75\,\mathrm{eV}$ & $\lambda_{0AB}^y$ & $0.011\,\mathrm{eV}$ \\
$\mu_d$ & $0.74\,\mathrm{eV}$ & $\lambda_{0}^y$ & $0.051\,\mathrm{eV}$ \\
$t_{px}$ & $1.13\,\mathrm{eV}$ & $\lambda_{0}^z$ & $0.012\,\mathrm{eV}$ \\
$t_{dx}$ & $-0.41\,\mathrm{eV}$ & $\lambda_{0}^{'y}$ & $0.050\,\mathrm{eV}$ \\
$t_{pAB}$ & $0.40\,\mathrm{eV}$ & $\lambda_{0}^{'z}$ & $0.012\,\mathrm{eV}$ \\
$t_{dAB}$ & $0.51\,\mathrm{eV}$  & $\lambda_{px}^{y}$ & $-0.040\,\mathrm{eV}$ \\
$t_{0AB}$ & $0.39\,\mathrm{eV}$ & $\lambda_{px}^{z}$ & $-0.010\,\mathrm{eV}$ \\
$t_{0ABx}$ & $0.29\,\mathrm{eV}$ & $\lambda_{dx}^{y}$ & $-0.031\,\mathrm{eV}$ \\
$t_{0x}$ & $0.14\,\mathrm{eV}$ & $\lambda_{dx}^{z}$ & $-0.008\,\mathrm{eV}$ \\
$t_{py}$ & $0.13\,\mathrm{eV}$ &&\\
\hline
$a$ & $3.477\,\mbox{\AA}$ & $b$ &  $6.249\,\mbox{\AA}$ \\
$\mathbf{r}_{Ad}$ & $(-0.25a, 0.32b)$ & $\mathbf{r}_{Bp}$ & $(0.25a, 0.07b)$ \\
$\mathbf{r}_{Ap}$ & $(-0.25a, -0.07b)$ & $\mathbf{r}_{Bd}$ & $(0.25a, -0.32b)$\\
\hline
\end{tabular}
\caption{Tight-binding and lattice parameters of the final 4-orbital model.}
\label{tab:tight_binding_parameters}
\end{table}
By making use of the Python package Qsymm~\cite{VRA18}, we generate all symmetry-allowed SOC terms up to nearest neighbors and also include symmetry-allowed, next-nearest neighbor SOC terms in the $x$ direction.
Figure~\ref{fig:WTe2_unit_cells}(c) illustrates all SOC terms taken into account.
The Bloch Hamiltonian of the final tight-binding model with SOC is $H(\mathbf{k})=s_0H_0(\mathbf{k})+ H_\mathrm{SOC}(\mathbf{k})$, with
\begin{eqnarray}
H_\mathrm{SOC} &=&
\big[(\lambda_{dx}^z\, s_z + \lambda_{dx}^y\, s_y)\sin(a k_x)\big]\, \Gamma_5^+ \nonumber\\
&& +\, \big[(\lambda_{px}^z\, s_z + \lambda_{px}^y\, s_y)\sin(a k_x)\big]\, \Gamma_5^- \nonumber\\
&& -i\lambda_{0AB}^{y}\, s_y\big(1 + e^{iak_x}\big)\,
e^{i\mathbf{k}\cdot\Delta_3}\, \Gamma_6  \nonumber\\
&&-i\big( \lambda_{0}^{z}\,s_z + \lambda_{0}^{y}\,s_y \big)
\Big( e^{i\mathbf{k}\cdot\Delta_4}\, \Gamma_4^+
- e^{-i\mathbf{k}\cdot\Delta_4}\, \Gamma_4^-\Big) \nonumber\\
&&-i\big( \lambda_{0}^{'z} s_z + \lambda_{0}^{'y} s_y \big) \nonumber\\
&&\hspace{1em}\times \Big( e^{-ibk_y} e^{i\mathbf{k}\cdot\Delta_4}\, \Gamma_4^+
- e^{ibk_y} e^{-i\mathbf{k}\cdot\Delta_4}\, \Gamma_4^-\Big) \nonumber\\
&& +\, \mathrm{H.c.},
\label{eq:SOC_terms}
\end{eqnarray}
where the $s_{x,y,z}$ are Pauli matrices acting in spin space and $s_0$ is the corresponding identity.

We obtain the parameters of the SOC terms in Eq.~\eqref{eq:SOC_terms} by fitting to ARPES data close to the Fermi level from Ref.~\onlinecite{TZW17}.
The large number of free parameters in the fit is handled using a LASSO regression analysis (see Supplemental Material~\cite{supp} for details).
As illustrated in Fig.~\ref{fig:SOC_fit}, the energy bands of the resulting model provide an excellent fit to the ARPES data. We have tabulated the parameter values of the final Hamiltonian in Tab.~\ref{tab:tight_binding_parameters}.

\begin{figure}[t]\centering
\includegraphics[width=1.0\columnwidth]
{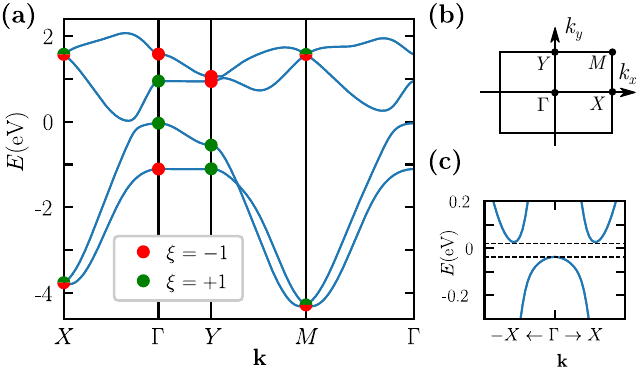}
\caption{Bulk energy bands of the monolayer-WTe$_2$ tight-binding model: (a) along high-symmetry lines in the Brillouin zone. We have indicated the parities $\xi$ of states at the time-reversal invariant momenta. (b) Bulk Brillouin zone. (c) Energy bands close to $\Gamma$ along $\Gamma X$ in a small energy window around the bulk energy gap. The dashed lines indicate the edges of the bulk energy gap.}
\label{fig:WTe2_bulk_bands}
\end{figure}
Figure~\ref{fig:WTe2_bulk_bands} shows the full band structure of the tight-binding model.
Due to the simultaneous presence of time-reversal and inversion symmetry, all bands are two-fold degenerate.
Most importantly, the model has an indirect energy gap of $\Delta E = 56\,\mathrm{meV}$, compatible with experiments~\cite{JSL17,TZW17,CGC18}.
Moreover, close to the Fermi level, the dispersion of our model matches experimental results~\cite{TZW17,CGC18} by construction.
An analysis of the parities $\xi$ of the states at the four time-reversal invariant momenta $\Gamma$, $X$, $Y$, and $M$ shows that the lowest conduction band and the lowest valence band in our model are inverted at $\Gamma$ [see Fig.~\ref{fig:WTe2_bulk_bands}(a)], as expected from DFT calculations~\cite{CSC16}. Finally, we compute the $\Z_2$ invariant of the model at half filling using the parities of occupied states at time-reversal invariant momenta~\cite{FuK07}. We obtain $\nu=1$, confirming that our model realizes a QSH insulator, in agreement with experiments~\cite{TZW17,WFG18}.

\section{Finite geometries without magnetic field}

With an accurate bulk tight-binding model at hand, we now study how its low-energy electronic properties depend on the specific termination in a finite geometry.
Obtaining an equally accurate model of the edge is not possible from the available experimental data.
Nevertheless, using a truncated bulk Hamiltonian is sufficient to understand the qualitative effects of various lattice terminations (the limitations of this approach are discussed in the Supplemental Material~\cite{supp}).
For that purpose, we put our monolayer-WTe$_2$ model into different ribbon geometries, each of which has only a single direction $\mathbf{d}$ of translational symmetry.
In this direction, we impose periodic boundary conditions.
In the perpendicular direction, in which the ribbon has a width $W$, we use open boundary conditions.
Here, we discuss four representative examples of terminations, which are depicted in the insets of Fig.~\ref{fig:edge_spectra}.
For these systems, we choose $W$ to be between $80$ and $100$ unit cells.
In the following, we present energy spectra and 2-terminal conductance calculations for these geometries, for now, without magnetic field.

\subsection{Energy spectra}

\begin{figure}[t]\centering
\includegraphics[width=1.0\columnwidth]
{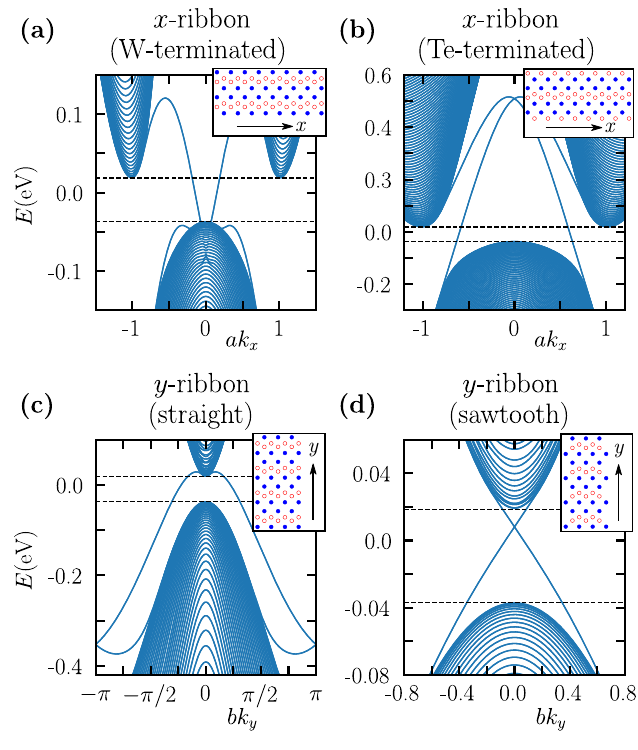}
\caption{Energy dispersion of ribbons with different terminations close to the Fermi level without magnetic field. The dashed lines indicate the bulk energy gap. The insets show the respective terminations using the same color code as in Fig.~\ref{fig:WTe2_unit_cells}(c).}
\label{fig:edge_spectra}
\end{figure}

As we have shown in the previous section, our model realizes a 2D topological insulator. Hence, for any finite geometry isolated bands of helical edge states are expected to connect the bulk valence and conduction bands across the bulk energy gap. To analyze how the dispersion of these edge bands depend on the considered termination, we compute the energy states of the ribbons as a function of the momentum $k_\parallel$ parallel to the translationally invariant direction $\mathbf{d}$.
Figure~\ref{fig:edge_spectra} shows the resulting energy spectra close to the Fermi level.

The general observations are the same for all ribbons: two pairs of isolated bands, corresponding to counterpropagating helical edge states, traverse the bulk energy gap.
Moreover, these states are spatially separated and localized to opposite edges of the ribbon.
This is a hallmark of QSH insulators.
The specific edge-state dispersions, however, differ qualitatively.

Figure~\ref{fig:edge_spectra}(a) shows the dispersion of a ribbon terminated between the W sites with $\mathbf{d}=\hat{\mathbf{x}}$.
This termination has edge energy bands that do not cross each other.
In contrast to that, ribbons terminated between Te sites with $\mathbf{d}=\hat{\mathbf{x}}$ have edge bands that cross at the time-reversal invariant edge momentum $k_\parallel =k_x=0$ [see Fig.~\ref{fig:edge_spectra}(b)].
The crossing represents a Kramers doublet, or \emph{Dirac point}, and is protected by time-reversal symmetry.
It is, however, energetically far outside the bulk energy gap.
We make a similar observation for straight terminations with $\mathbf{d}=\hat{\mathbf{y}}$, as shown in Fig.~\ref{fig:edge_spectra}(c).
Here, the Dirac point is at $k_\parallel =k_y=\pi/b$ and energetically far below the bulk energy gap.

The situation is different for ribbons with $\mathbf{d}=\hat{\mathbf{y}}$ and a sawtooth termination [see Fig.~\ref{fig:edge_spectra}(d)].
Again, we find two pairs of helical edge bands traversing the bulk energy gap, but in this geometry the bands cross inside the bulk energy gap at the time-reversal invariant edge momentum $k_\parallel =k_y=0$: the edge-state spectrum has a so-called \emph{in-gap} Dirac point.
In contrast, the edge Dirac points of most terminations, including the ones discussed before, are ``hidden'' or \emph{buried}~\cite{SPA18} in the bulk energy bands.
This is due to the strong electron-hole asymmetry of the material.
We will see that the distinction between terminations with buried and with in-gap edge Dirac points is crucial in the presence of a magnetic field.

\begin{figure}[t]\centering
\includegraphics[width=1.0\columnwidth]
{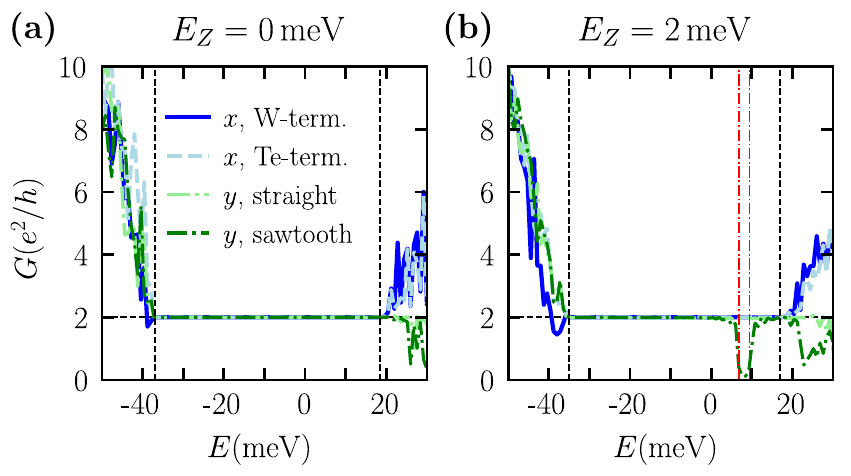}
\caption{2-terminal conductance of differently terminated WTe$_2$ ribbons: (a) without magnetic field ($E_Z=0$); (b) with an out-of-plane magnetic field corresponding to $E_Z=2\,\mathrm{meV}$ ($B=9.4\,\mathrm{T}$). The dashed vertical lines indicate the edges of the bulk conduction and of the bulk valence band. We have also indicated the induced edge energy gap of the sawtooth $y$-ribbon by vertical dash-dot lines in red.}
\label{fig:conductance_regular}
\end{figure}

\subsection{2-terminal conductance}

We now look at how the termination of the ribbons affects their transport properties.
Specifically, we look into the 2-terminal conductance of our systems. For that purpose, we remove the periodic boundary conditions of the ribbon under consideration and attach metallic leads of the same width as the ribbon on each side. 
This is achieved by shifting the chemical potential deep into the valence band at both ends of the ribbon.
The samples considered here have a length $L$ of 200 unit cells.
This is of similar magnitude as the edge length of samples studied in experiments with sizes ranging from 60 to 100 nm\cite{WFG18}.

We compute the corresponding scattering matrix $S(E)$ as a function of energy $E$ using the quantum transport software package Kwant~\cite{GWA14}.
The conductance is then given by $G = e^2/h \sum_{nm}|S_{nm}|^2$, where $n,m$ run over all electronic modes in the leads.
We show our results for $G$ as a function of $E$ without magnetic field in Fig.~\ref{fig:conductance_regular}(a).

\section{Finite geometries with magnetic field}

We proceed by studying the effect of an out-of-plane magnetic field $\mathbf{B}=B\hat{\mathbf{z}}$ on the edge-state dispersions and on the conductance of monolayer WTe$_2$.
For this purpose, we add an on-site Zeeman term of the form $H_Z = E_Z\,s_z\Gamma_0$ to the time-reversal symmetric Hamiltonian $H(\mathbf{k})$ of Eqs.~\eqref{eq:spinless_Hamiltonian} and~\eqref{eq:SOC_terms}, where $\Gamma_0$ is the $4\times 4$ identity matrix in orbital space.
We establish the relation between the Zeeman energy $E_Z$ and the magnetic field $B$ later on.

\begin{figure}[t]\centering
\includegraphics[width=1.0\columnwidth]
{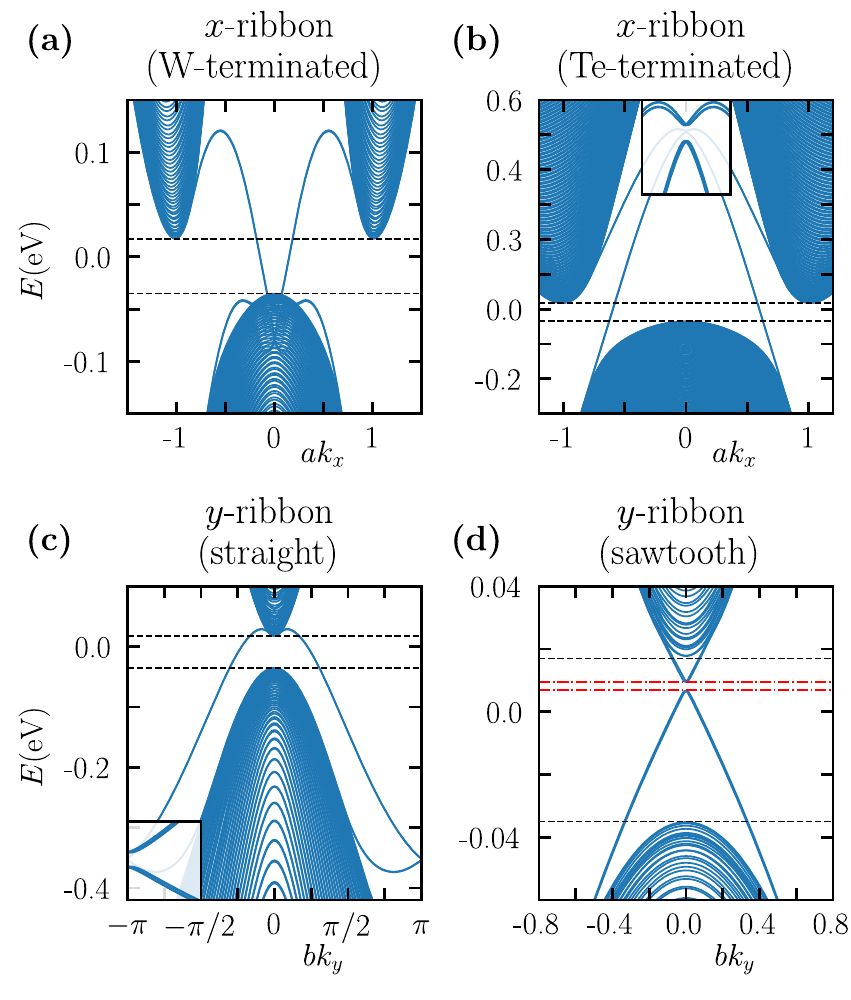}
\caption{Energy dispersion of ribbons with different terminations close to the Fermi level with an out-of-plane magnetic field corresponding to $E_Z=2\,\mathrm{meV}$ ($B=9.4\,\mathrm{T}$). The dashed lines indicate the edges of the bulk energy gap. We have also highlighted the edge energy gap of the sawtooth $y$-ribbon by red dash-dot lines.
The insets in panels (b) and (c) show magnified spectra around the gapped edge Dirac points.}
\label{fig:edge_spectra_magnetic}
\end{figure}

Figure~\ref{fig:edge_spectra_magnetic} shows energy spectra corresponding to the ribbons studied in the previous section under a finite magnetic field.
First of all, we observe that the bulk energy gap is slightly smaller.
Moreover, we generally find that the degeneracy of edge states localized to \emph{opposite} boundaries of the ribbon is lifted because of the broken time-reversal symmetry.
However, a Zeeman gap opens in the edge-state spectrum only if the lifted degeneracy involves states localized to the \emph{same} boundary, such as states of a Dirac point.
We discuss the implications below.

W-terminated $x$-ribbons do not have a Dirac point in their edge-state spectrum.
Consequently, the edge-state dispersion remains gapless even at finite magnetic fields, as we show in Fig.~\ref{fig:edge_spectra_magnetic}(a).
In contrast to that, the spectra of Te-terminated $x$-ribbons and of straight $y$-ribbons do have an edge Dirac point far outside the bulk energy gap.
These Dirac points are gapped out by the magnetic field, but lead only to a ``warped'' Zeeman gap in the energy spectrum.
In other words, bulk valence and conduction bands are no longer connected by edge bands, but there still exist electronic states (bulk and/or edge states) for any fixed energy $E$ [see Figs.~\ref{fig:edge_spectra_magnetic}(b)-(c)].
Due to the nature of the edge energy gap, the ribbons are effectively still gapless  in the presence of a magnetic field.
Hence, the edge Zeeman gap, without other states crossing it, must be \emph{inside} the bulk energy gap to suppress electronic transport. 

The sawtooth $y$-termination satisfies this requirement with an edge Zeeman gap at $k_y=0$ inside the bulk energy gap, which we show in Fig.~\ref{fig:edge_spectra_magnetic}(d). This behavior is fundamentally different from the other ribbon geometries: there now exists a small energy window inside the bulk energy gap without any available states.
Moreover, from the size of the Zeeman gap and from the effective out-of-plane $g$ factor~\cite{WFG18}, we estimate the relation between the Zeemann energy $E_Z$ and the magnetic field strength $B$ to be $B[\mathrm{T}] \approx 4.7\,E_Z[\mathrm{meV}]$ (see Supplemental Material~\cite{supp} for details).

Also the conductance reflects the qualitative difference of edge-state spectra between different ribbons.
Figure~\ref{fig:conductance_regular}(b) shows the conductance corresponding to the ribbons analyzed above as a function of energy under magnetic field. We still find $G=2e^2/h$ throughout the bulk energy gap for the two $x$-ribbons and for the straight $y$-ribbon.
The conductance remains quantized because the considered edges preserve translational symmetry.
On the contrary, the conductance of the sawtooth $y$-ribbon drops to zero around the Dirac-point energy, while being quantized to $2e^2/h$ outside this small energy window.
Furthermore, by comparing to the edge-state spectrum in Fig.~\ref{fig:edge_spectra_magnetic}(d), we directly attribute the position and the width of the conductance drop to the presence of an edge energy gap.

\section{Ribbons with disordered edges under magnetic field}

The edges of typical samples used in experiments on 2D monolayer materials are highly irregular, i.e., they have cracks, steps or bumps~\cite{WFG18,CGC18}.
Hence, the samples are no longer translationally invariant along their boundary, enabling backscattering of edge states when time-reversal symmetry is broken.
\begin{figure}[t]\centering
\includegraphics[width=1.0\columnwidth]
{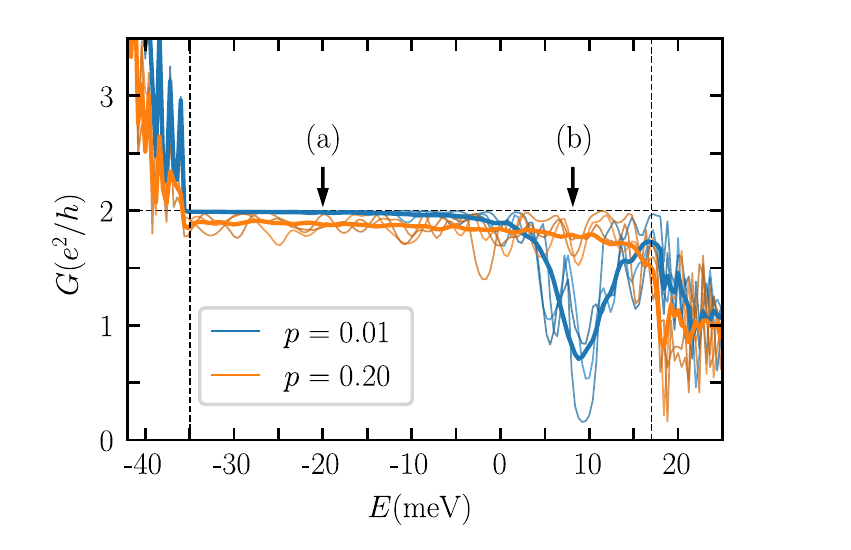}
\caption{Conductance $G$ of sawtooth $y$-ribbons under magnetic field [$E_Z=2\,\mathrm{meV}$ ($B=9.4\,\mathrm{T}$)] as a function of energy $E$ for different disorder strengths $p$. Thin lines represent single samples with different realizations of disordered edges at fixed $p$. Bold lines correspond to the average conductance $G_{av}$ for an ensemble of $50$ disorder realizations. The vertical dashed lines indicate the edges of the bulk energy gap. The black arrows point to the energy values considered in Figs.~\ref{fig:magnetoconductance_y_ribbon}(a) and~(b), respectively.}
\label{fig:conductance_magnetic_disordered}
\end{figure}
We model a rough edge by the boundary following a random walk with a step appearing at every site with probability $p$ (see Supplemental Material~\cite{supp} for details).
The parameter $p$ is therefore a measure of disorder strength.

Figure~\ref{fig:conductance_magnetic_disordered} shows the conductance $G$ of single sawtooth $y$-ribbons and the corresponding disorder-averaged conductance $G_{av}$ for different disorder strengths $p$.
For a given disordered sample the conductance fluctuates considerably as a function of energy.
We find that the characteristic exponential suppression of conductance is visible only in the low-disorder regime for edges with an in-gap Dirac point.
This is also reflected in a dip of the average conductance $G_{av}$ around the Dirac-point energy.
Away from a Dirac point, however, $G_{av}$ shows only a weak energy dependence.
The height of this conductance plateau decreases with increasing disorder $p$ due to the onset of Anderson localization.
\begin{figure}[t]\centering
\includegraphics[width=1.0\columnwidth]
{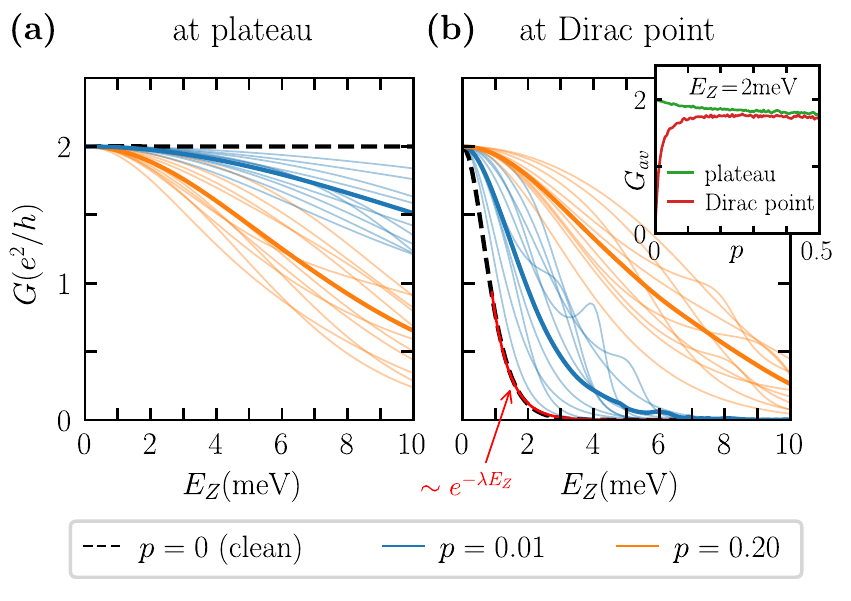}
\caption{Conductance $G$ of sawtooth $y$-ribbons at fixed energies $E$ as a function of the disorder strength $p$ and the Zeeman energy $E_Z$:
(a) at the plateau ($E=-20\,\mathrm{meV}$);
(b) at the Dirac point ($E=8.15\,\mathrm{meV}$) with exponential decay in the clean limit ($\lambda=1.8\,\mathrm{meV}^{-1}$).
Thin lines correspond to single samples with particular realizations of edge disorder at fixed $p$.
Bold lines depict the average conductance $G_{av}$ over $50$ disorder realizations.
The inset in panel (b) shows the average conductance as a function of $p$ at fixed $E_Z=2\,\mathrm{meV}$ ($B=9.4\,\mathrm{T}$).}
\label{fig:magnetoconductance_y_ribbon}
\end{figure}
On the other hand, strong disorder removes the conductance suppression in the Zeeman gap and the conductance shows the same plateau behavior throughout the entire bulk energy gap.
In other words, the edge energy gap effectively closes for strong disorder, thereby becoming statistically indistinguishable from an edge without in-gap Dirac point [see Fig.~\ref{fig:conductance_magnetic_disordered}].

We observe a qualitatively similar behavior for the conductance as a function of the Zeeman energy $E_Z$ (see Fig.~\ref{fig:magnetoconductance_y_ribbon}).
For weak disorder, the conductance at the Dirac point decays exponentially to zero, which is in agreement with experimental results~\cite{WFG18}, whereas it decays much slower in the plateau regime.
As expected from Anderson localization, the conductance in the plateau regime is suppressed with increasing disorder.
Contrary to this, the conductance at the Dirac point is \emph{enhanced} by the edge disorder.
This happens because localized states, whose finite overlap with the leads enables transport, fill the Zeeman gap.
Moreover, as we show in the inset of Fig.~\ref{fig:magnetoconductance_y_ribbon}(b), the $G_{av}(p)$ curves of the two regimes gradually approach each other until they coincide for sufficiently strong disorder.

We have seen that, in the clean limit, there are two qualitatively different behaviors in the presence of magnetic field.
For samples with an in-gap edge Dirac point, the conductance drops to zero around the energy of the Dirac point.
For terminations with a buried Dirac point, the conductance is constant and equal to $2e^2/h$.
We observe the same plateau behavior of quantized conductance also for terminations with in-gap Dirac points if the conductance is measured away from the Dirac-point energy.
While this is qualitatively reproduced in samples with weak edge disorder, strong disorder washes out the Zeeman gap.
The experiments in Ref.~\onlinecite{WFG18} have observed exponential conductance suppression.
Therefore, we expect the edges of the samples investigated there to contain sufficiently long straight segments.

\section{Conclusion}

We have studied edge-state dispersions and 2-terminal conductance of the quantum spin-Hall insulator monolayer $1T'$-WTe$_2$ in a 4-orbital tight-binding model in various geometries. We have derived our model combining density-functional theory calculations, symmetry considerations, and photo-emission spectroscopy data.
By construction, our model provides a better fit to experimental results than previous approaches.
We use this model to study the effects of magnetic field and disorder on differently terminated nanoribbons.
Without magnetic field, the topological nature of the system gives rise to helical edge states independent of the system geometry. Nonetheless, we find that the edge-state dispersion strongly depends on the termination: for some terminations there is an in-gap edge Dirac point, whereas for others the edge Dirac point is buried in the bulk energy bands. This has important consequences for the conductance in the time-reversal symmetry broken regime.

For terminations with a buried edge Dirac point, the conductance fluctuates around a plateau, the value of which decreases slowly with magnetic field and with the magnitude of disorder along the edge. For terminations with an in-gap Dirac point, there is an additional suppression of conductance around the energy of the Dirac point due to a Zeeman gap. There, the conductance exponentially decays to zero as a function of the magnetic field. We further observe that the conductance is gradually enhanced around the Dirac-point energy as the edge disorder is increased. Hence, the characteristic exponential suppression of conductance is only visible for sufficiently clean edge terminations with an in-gap Dirac point.

Our results help to understand recent experimental findings in $1T'$-WTe$_2$ monolayers.
Moreover, the tight-binding model derived in this work provides a minimal but realistic low-energy description to study other promising directions in this materials class, such as the $1T'$-phase~\cite{UPT18,CPC18} in WSe$_2$ or superconductivity in $1T'$-WTe$_2$ monolayers~\cite{FWC18,SPF18}.

\begin{acknowledgments}
We thank L. Wang, V. Fatemi, M. Wimmer, M. Richter, K. Koepernik, and J. Facio for fruitful discussions and helpful comments.
We thank S. Tang and Z.-X. Shen for providing access to their experimental data.
R. Ray thanks U. Nitzsche for technical support.
This work was supported by ERC Starting Grant 638760, the Netherlands Organisation for Scientific Research (NWO/OCW), and the US Office of Naval Research.
R. Ray acknowledges financial support from the European Union (ERDF) and the Free State of Saxony via the ESF projects 100231947 and 100339533 (Young Investigators Group “Computer Simulations for Materials Design” – CoSiMa).

\emph{Author contributions:} A. Akhmerov initiated and oversaw the project.
A. Lau and D. Varjas formulated the minimal model from symmetries.
R. Ray carried out the density-functional theory calculations.
A. Lau and R. Ray performed the fitting of the tight-binding parameters.
A. Lau performed the conductivity and energy spectra calculations for the tight-binding model.
D. Varjas took part in modeling the disordered edges.
A. Lau, D. Varjas, R. Ray, and A. Akhmerov contributed to interpreting the results and writing the manuscript.

\emph{Data availability:} All files and data used in this study are available in the repository at Ref.~\onlinecite{repo}.

\end{acknowledgments}



\newpage
\clearpage
\appendix

\renewcommand\thefigure{S.\arabic{figure}}    
\renewcommand\theHfigure{S.\arabic{figure}} 
\setcounter{figure}{0}   

\onecolumngrid
\section*{SUPPLEMENTAL MATERIAL}

\twocolumngrid

\section{Details of model construction}

In the main text, we construct a minimal 4-orbital tight-binding model for monolayer WTe$_2$. In the following, we provide more details on the single steps of the construction.

\subsection{DFT calculations and optimized Wannier-orbital fit}

First of all, to capture the essential features of the scalar-relativistic density-functional theory (DFT) bandstructure at low energies, we have to include a large number of neighbors in the Wannier-orbital fit (hopping range up to $20$ {\AA}).
Irrespective of the number of neighbors in the low-energy model, however, the location of the Dirac point in $\mathbf{k}$ space is slightly shifted compared to the DFT band structure.
While a more complex 6-orbital model does not suffer from such discrepancies, also within the minimal 4-orbital model we obtain an accurate low-energy description by adopting the hybrid approach described below.

First, we systematically reduce the number of neighbors by introducing a cut-off, keeping only hopping terms with magnitude larger than or equal to $0.06\,\mathrm{eV}$. In addition, we include second-nearest neighbor hopping $t_{py}$ in the $y$-direction, which is slightly smaller than the cuf-off.
This is necessary to avoid a nearly flat dispersion of the highest valence band along the $\Gamma$--$Y$ direction, since $t_{py}$ has a dominant contribution to the finite curvature along this line.
Next, we fix the onsite energies and all dominant hopping parameters $\geq 0.40\,\mathrm{eV}$.
We treat the remaining parameters as effective to account for neglected longer-range hopping terms and for the discrepancies between the Wannier model and the band structure mentioned above.

We then fit the effective parameters to the DFT energy bands close to the Fermi level to capture accurately the low-energy details of the band structure.
To keep the deviations small and to reduce the number of model parameters further, we employ a LASSO regression analysis with a regularization parameter $\lambda$.
We determine the optimal $\lambda$ such that the modified model parameters are close to their original values, with the largest change equal to $0.17\,\mathrm{eV}$, and such that they are all smaller than the dominant hopping parameters.
Finally, we discard the parameters smaller than the cut-off value.

\subsection{Fitting SOC parameters to ARPES data}

\begin{figure}[t]\centering
\includegraphics[width=1.0\columnwidth]
{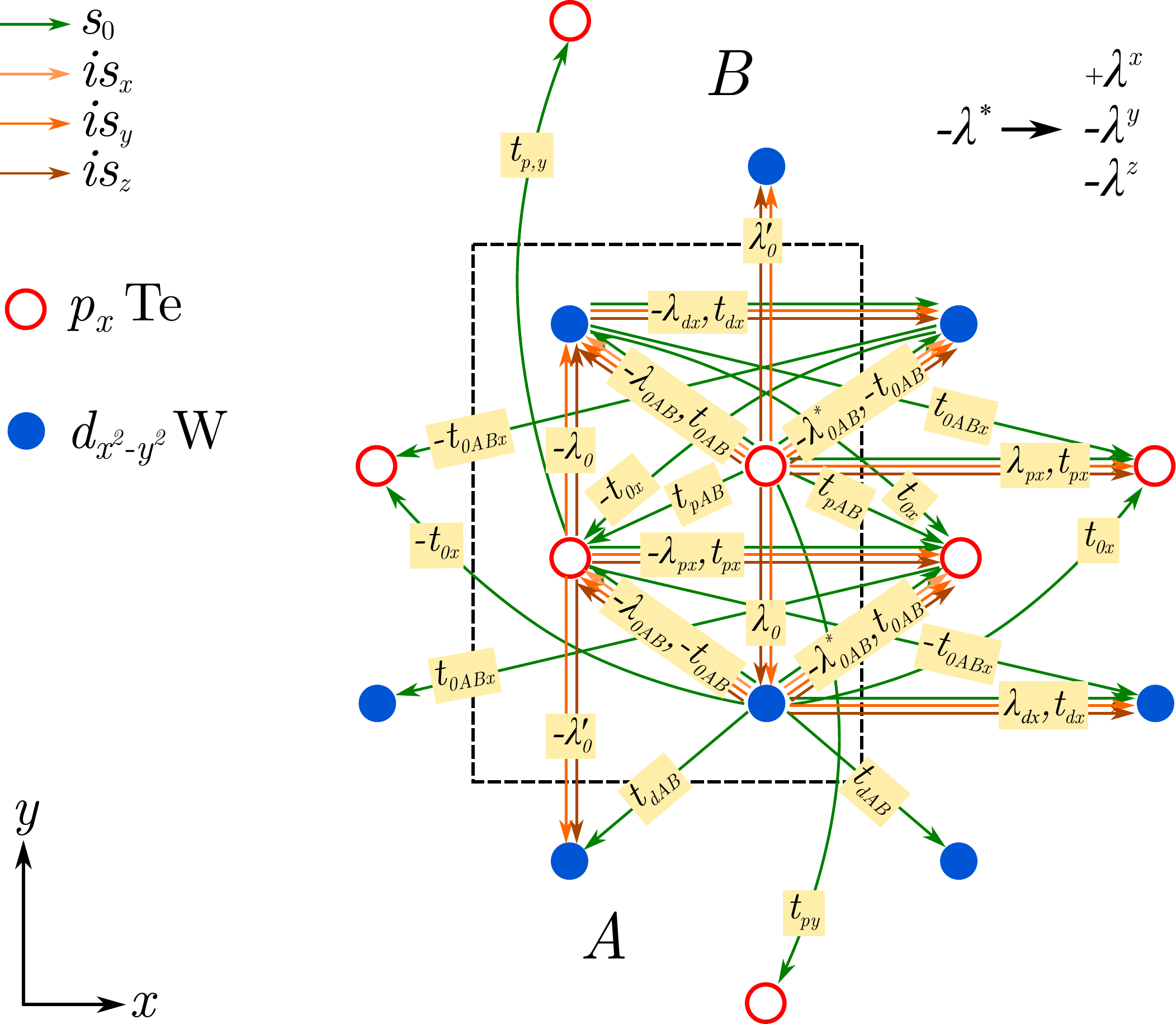}
\caption{Cartoon of hopping and symmetry-allowed SOC terms.}
\label{fig:WTe2_tb_parameters}
\end{figure}
We now add all symmetry-allowed spin-orbit coupling (SOC) terms up to nearest neighbors to our model.
Furthermore, we also add symmetry-allowed, next-nearest neighbor SOC terms in the $x$ direction.
We expect those to be non-negligible since the parameters of the related hopping terms $t_{px}$ and $t_{dx}$ are among the largest in the spinless Hamiltonian (see Tab.~1 in the main text).
The SOC terms are illustrated in Fig.~\ref{fig:WTe2_tb_parameters}.
To determine the corresponding SOC parameters, we choose to fit these parameters to experimental data from angle-resolved photoemission spectroscopy (ARPES) close to the Fermi level.
The raw data has been provided by the authors of Ref.~\onlinecite{TZW17}.

\begin{figure}[t]\centering
\includegraphics[width=1.0\columnwidth]
{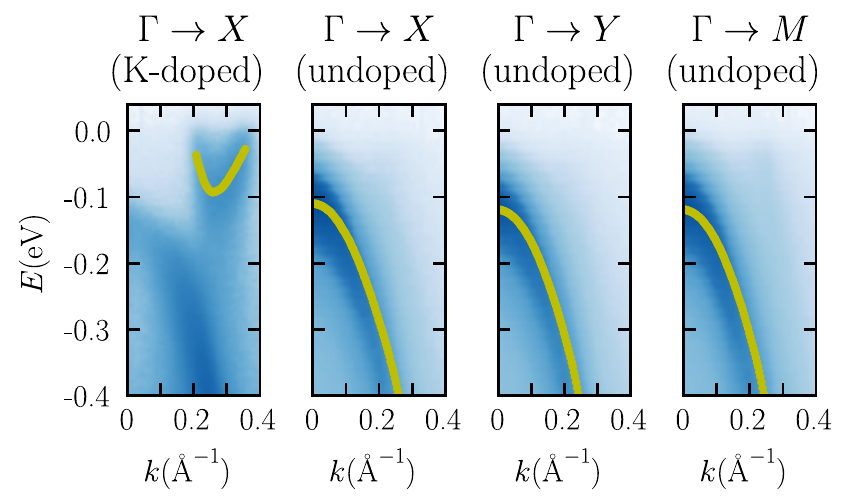}
\caption{ARPES spectra (blue) from Ref.~\onlinecite{TZW17} close to the Fermi level and overlaid intensity peaks (yellow).}
\label{fig:ARPES_raw_peaks}
\end{figure}

The raw data comprises measured ARPES spectra of monolayer WTe$_2$ along three different directions through the two-dimensional Brillouin zone, namely along $\Gamma X$, along $\Gamma Y$, and along $\Gamma M$.
Furthermore, we also use ARPES raw data for K-doped monolayer WTe$_2$ along $\Gamma X$, for which the Fermi level lies in the conduction band.
From the raw data we determine the peaks of the spectral weights by fitting to ensembles of Lorentzians.
In this way, we obtain experimental energy bands for the first valence band close to $\Gamma$ along three different directions, and for a small piece of the first conduction band along $\Gamma X$ (see Fig.~\ref{fig:ARPES_raw_peaks}).
The distance in energy between the valence and the conduction bands is set to the value of $56\,\mathrm{meV}$, obtained from scanning tunneling spectroscopy (STS) measurements on the same samples in Ref.~\onlinecite{TZW17}.

Besides a good agreement with the ARPES peaks for the individual energy bands, we require the induced indirect energy gap to equal the STS value.
To account for discrepancies between ARPES data and DFT, we further allow for small perturbations of the two smallest hopping parameters $t_{py}$ and $t_{0x}$, which we obtained above for the tight-binding model without SOC.
To handle the large number of free parameters, we again employ a LASSO regression analysis.
We choose the optimal regularization parameter such that the model has an energy gap identical to the STS value and such that the perturbations of the spinless parameters are $\leq 0.02\,\mathrm{eV}$.
Finally, we discard SOC parameters $<8\,\mathrm{meV}$, which does not affect the quality of the fit result considerably.

\subsection{The final tight-binding model}

The schematic in Fig.~\ref{fig:WTe2_tb_parameters} illustrates the hopping and SOC terms with their corresponding parameters.
Based on this, the full tight-binding model given in the main text can be assembled by making use of translational symmetry and Hermiticity only.
Relations between terms connected by other symmetries are captured by the notation.

In the main text, we also provide the corresponding Bloch Hamiltonian $H(\mathbf{k})=H_0(\mathbf{k}) + H_{SOC}(\mathbf{k})$.
There, we introduce the $4\times 4$ matrices $\Gamma_i$ defined by:
\begin{eqnarray}
\Gamma_0 &=& \tau_0\sigma_0,\\
\Gamma_1^\pm &=& \frac{\tau_0}{2}(\sigma_0 \pm \sigma_3),\\
\Gamma_2^\pm &=& \frac{1}{4}(\tau_1 + i\tau_2)(\sigma_0 \pm \sigma_3),\\
\Gamma_3 &=& \frac{1}{2}(\tau_1 + i\tau_2)i\sigma_2, \\
\Gamma_4^\pm &=& \frac{1}{4}(\tau_0 \pm \tau_3)(\sigma_1 + i\sigma_2),\\
\Gamma_5^\pm &=& \frac{\tau_3}{2}(\sigma_0 \pm \sigma_3),\\
\Gamma_6 &=& \frac{1}{2}(\tau_1 + i\tau_2)\sigma_1,
\end{eqnarray}
where the matrices $\tau_j\sigma_i$ are products of Pauli matrices acting in orbital space with respect to the basis $\lbrace d_{\mathbf{k}Ads}, d_{\mathbf{k}Aps}, d_{\mathbf{k}Bds}, d_{\mathbf{k}Bps}\rbrace$, where $d_{\mathbf{k}cls}$ annihilates an electron with momentum $\mathbf{k}$, spin-$S_z$ eigenvalue $s=\uparrow,\downarrow$ and orbital $l=p,d$ (Te,W) in sublattice $c=A,B$.
More specifically, $\tau_j$ acts on the sublattice degree of freedom and $\sigma_i$ acts on the orbital degree of freedom.

\begin{figure}[t]\centering
\includegraphics[width=1.0\columnwidth]
{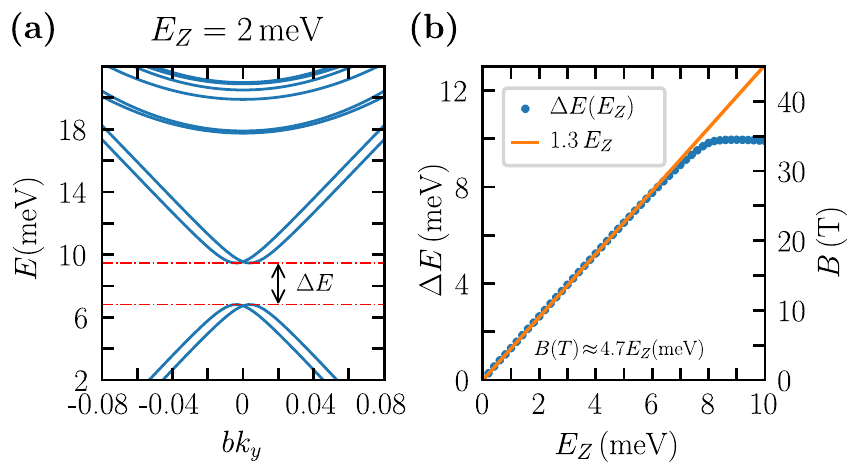}
\caption{Edge energy gap for sawtooth $y$-ribbons under magnetic field: (a) magnified ribbon spectrum around the edge energy gap $\Delta E$, which we have indicated by red, dash-dot lines. (b) Edge energy gap $\Delta E$ and magnetic field $B$ as a function of the Zeeman energy $E_Z$.}
\label{fig:magnetic_field}
\end{figure}

\section{Determining the magnetic field from the edge gap}

In the main text, we study the effect of an onsite Zeeman term in the Hamiltonian of monolayer WTe$_2$ on the edge dispersion of differently terminated ribbons. Since it breaks time-reversal symmetry, it gaps out the edge Dirac points in the ribbon spectrum, if present. If the edge Dirac point lies inside the energy gap of the bulk bands, the Zeeman term thus opens a direct energy gap in the spectrum of the ribbon. In particular, we have seen that $y$-ribbons with a sawtooth termination realize this edge Zeeman gap. Figure~\ref{fig:magnetic_field}(a) shows a magnified region of the ribbon spectrum around the edge gap.

We now use the size of the observed edge Zeeman gap $\Delta E$ to determine the relation between the magnetic field $\mathbf{B}=B\mathbf{e}_z$ and the Zeeman energy $E_Z$. For that purpose, we make use of the relation
\begin{equation}
\Delta E = g\mu_B B,
\label{B_dE_relation}
\end{equation}
where $\mu_B$ is the Bohr magneton and $g$ is the effective out-of-plane $g$ factor. A recent experiment~\cite{WFG18} has observed an edge Zeeman gap in WTe$_2$ monolayers and has obtained a value of $g=4.8$.

We have computed ribbon spectra for a number of values for the Zeeman energy $E_Z$ and determined the size of the induced Zeeman gap $\Delta E$ for each realization [see Fig.~\ref{fig:magnetic_field}(a)].
We plot the results in Fig~\ref{fig:magnetic_field}(b), where we have also added a second axis to indicate the corresponding magnetic field $B$ based on Eq.~\eqref{B_dE_relation} and on the experimental $g$ factor. We find that the relation between $E_Z$ and $B$ is approximately linear for small values of $E_Z$. More specifically, we determine the linear relation to be
\begin{equation}
B[\mathrm{T}] = 4.7\,E_Z[\mathrm{meV}].
\end{equation}
At large magnetic fields, the growing edge gap is compensated by the shrinking bulk gap of the system.

\begin{figure}[t]\centering
\includegraphics[width=1.0\columnwidth]
{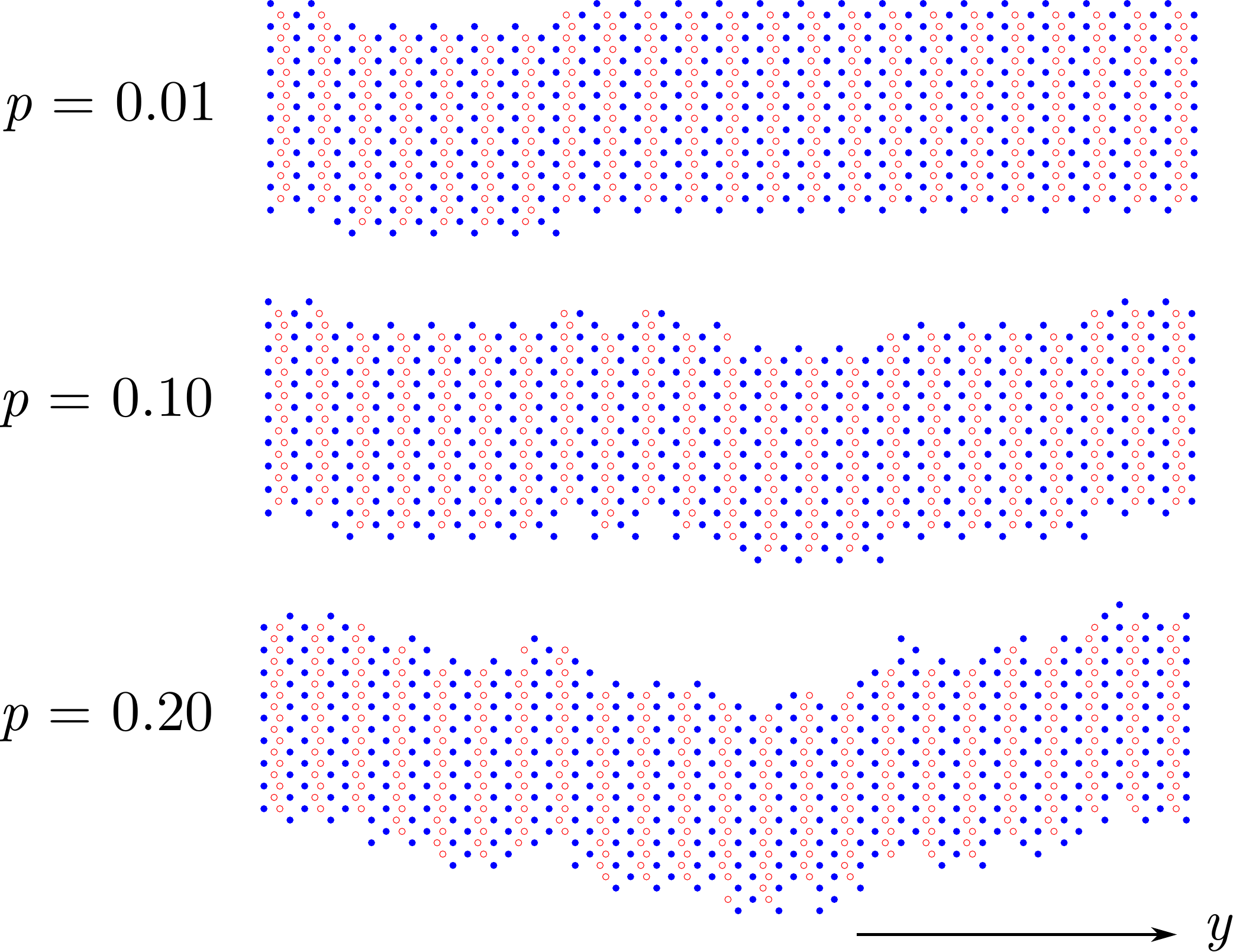}
\caption{Geometries of sawtooth $y$-ribbons with disordered edge terminations as a function of the disorder parameter $p$.}
\label{fig:irregular_ribbons}
\end{figure}

\section{Modeling of disordered ribbons}

We model disordered (or rough) edges by adding and removing lattice sites along an initial, translationally invariant boundary following a random walk $x_p(y)$ with parameter $p$ and step widths $\Delta x$ and $\Delta y$. More specifically, a step by $\Delta y$ in the variable $y$ is accompanied by either a step of $+\Delta x$ in the variable $x$ with probability $p/2$, a step of $-\Delta x$ in $x$ with probability $p/2$, or no step in $x$ with probability $1-p$.

Starting from translationally invariant $y$-ribbons with sawtooth terminations, as done in the main text, we choose the step lengths in the following way: a step $\Delta y$ in the random walk corresponds to a step by one lattice site parallelly along the sawteeth, whereas a step $\Delta x$ in the random walk converts to a step by one lattice site in the $x$ direction. In particular, we choose $\Delta y = b/4$ and $\Delta x = a$. Furthermore, without loss of generality, we choose the initial and the final point of the random walk to be the same, i.e., $x_p(y_0)=x_p(y_0+N\Delta y)$, where $N$ is the length of the random walk.

In Fig.~\ref{fig:irregular_ribbons} we show realizations of ribbons, subject to the prescription above, for different values of the disorder strength $p$. For small $p$, the edge is very unlikely to deviate from the initial sawtooth termination and we see long, non-disordered segments separated by steps. For larger $p$, however, deviations from the initial, clean edge become dominant.

\section{Limitations of the nanoribbon models}

In the main text, we model finite nanoribbon geometries by truncating the real-space bulk Hamiltonian.
As a reasonable starting point, this is expected to give a qualitative picture of the effects of such lattice terminations but comes with limitations.

In general, the edge resulting from the termination of a material is chemically active due to the presence of dangling bonds.
Depending on the experimental setup, this will lead to passivation with other atoms, such as oxygen or hydrogen.
In WTe$_2$ monolayers, which are encapsulated in boron nitride~\cite{WFG18}, such passivation is expected to involve boron and nitride.
However, the detailed effects of such an environment on the edges of WTe$_2$ monolayers are not known. 

\begin{figure}[t]\centering
\includegraphics[width=1.0\columnwidth]
{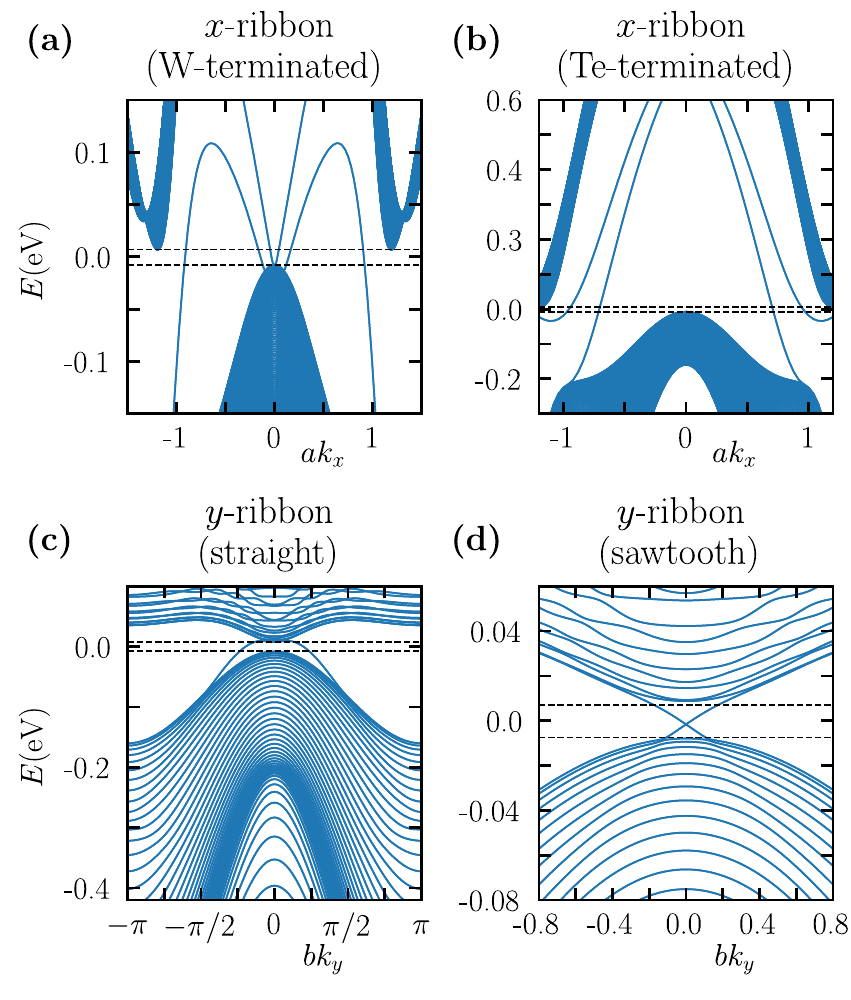}
\caption{Edge spectra with original Wannier bulk parameters without magnetic field for the same ribbon geometries as shown in the main text.
The dashed lines indicate the edges of the bulk valence and bulk conduction bands, respectively.}
\label{fig:bare_hopping_spectra}
\end{figure}
On the level of a low-energy model, such effects will mainly induce effective edge potentials that deform the edge-energy bands associated with a specific termination.
In particular, such potentials can shift the edge Dirac points energetically.
If the edge potentials are strong enough, this can push an edge Dirac point out of the bulk-energy gap or vice versa.
Moreover, edge potentials can lead to additional trivial edge bands lurking into the bulk-energy gap.
If the edge potentials are small, they will have only a perturbative effect on the edge dispersions obtained from the truncated bulk model.
Nonetheless, all of the effects above can be easily incorporated into the truncated bulk model once the induced edge potentials are known from experiments or from sufficiently accurate \textit{ab-initio} studies of nanoribbons.

\section{Effect of bulk-parameter perturbations on the edge dispersions}

In the main text, we optimize the spinless hopping parameters to obtain a low-energy bulk model with an accurate dispersion close to the Fermi level.
This results in a small deviation of the original Wannier parameters obtained from an optimization of the orbital overlaps.
In the following, we are going to show that this deviation has only a small quantitative effect on the edge dispersion of corresponding nanoribbons.

For that, we use all Wannier parameters with a magnitude larger than or equal to $0.06\,\mathrm{eV}$ plus the original second-nearest neighbor hopping $t_{py}$ in the $y$ direction. 
Furthermore, we apply the SOC parameters of the final model.
The edge spectra of nanoribbons with this set of bulk parameters are shown in Fig.~\ref{fig:bare_hopping_spectra}, where the panels correspond to the same terminations as in the main text.

First of all, we find that the bulk-energy gap, induced by the SOC terms, is much smaller than in the optimized model, namely $14\,\mathrm{meV}$ instead of $56\,\mathrm{meV}$.
This is due to the large deviation of the spinless Wannier model from the DFT energy bands close to the Fermi level, which underlines the need for optimization.
More specifically, the bulk Dirac point is much lower in energy with respect to the band maximum at $\Gamma$, while being at a momentum farther away from $\Gamma$ than in DFT (see Fig.~2 in the main text).
This discrepancy would require the SOC terms to be unphysically scaled up (by a factor of $1.3$) to achieve the experimentally observed bulk-energy gap.
In other words, the SOC gap at the Dirac points would be largely overestimated without optimizing the spinless bands close to the Fermi level.

Similar to the bulk bands, the edge bands are deformed by this parameter perturbation.
Nevertheless, their qualitative features remain the same: only the saw-tooth termination in Fig.~\ref{fig:bare_hopping_spectra}(d) features an in-gap edge Dirac point, whereas the Dirac points are buried for the other terminations.
Furthermore, also the relative location of the buried Dirac points is unchanged, i.e., whether they are buried in the bulk valence or bulk conduction band.
This suggests that these features are robust towards mild perturbations of the tight-binding parameters.

\end{document}